# Ultrathin BIC metasurfaces based on ultralow-loss Sb$_2$Se$_3$ phase-change material

Zhaoyang Xie[1], Chi Li[1,*], Krishna Murali[2], Haoyi Yu[1], Changxu Liu[3], Yiqing Lu[4],

Stefan A. Maier[1,5,*], Madhu Bhaskaran[2,*], Haoran Ren[1,*]

[1]School of Physics and Astronomy, Faculty of Science, Monash University, Melbourne, Victoria 3800, Australia.

[2]ARC Centre of Excellence for Transformative Meta-optical Systems (TMOS), Functional Materials and Microsystems Research Group, RMIT University, Melbourne, Victoria, 3000, Australia.

[3]Centre for Metamaterial Research & Innovation, Department of Engineering, University of Exeter, Exeter EX44QF, United Kingdom.

[4]School of Engineering, Macquarie University, Sydney, New South Wales 2109, Australia.

[5]Department of Physics, Imperial College London, London SW7 2AZ, United Kingdom.

* Corresponding authors:

Chi.Li1@monash.edu

Stefan.Maier@monash.edu

Madhu.Bhaskaran@rmit.edu.au

Haoran.Ren@monash.edu

**Abstract**

Phase-change materials (PCMs) are increasingly recognised as promising platforms for tunable photonic devices due to their ability to modulate optical properties through solid-state phase transitions. Ultrathin and low-loss PCMs are highly valued for their fast and more effective phase transitions and applications in reconfigurable photonic chips, metasurfaces, optical modulators, sensors, photonic memories, and neuromorphic computing. However, conventional PCMs such as GST, GSST, $VO_2$, and $In_3SbTe_2$, despite optimisation for tunable meta-optics, suffer from high intrinsic losses in the near-infrared (NIR) region, limiting their potential for high quality factor (Q-factor) resonant metasurfaces. Here we present the design and fabrication of tunable bound states in the continuum (BIC) metasurfaces using the ultralow-loss PCM $Sb_2Se_3$. Our BIC metasurfaces, only 25 nm thick, achieve high modulation depth and broad resonance tuning in the NIR with high Q-factors up to 130, without the need for additional materials. Experimentally, we employ these BIC metasurfaces to modulate photoluminescence in rare earth-doped upconversion nanoparticles, reducing the excitation power for multiphoton photoluminescence and enabling emission polarisation manipulation. This work offers a promising platform for developing active resonant metasurfaces in the NIR region, with broad applications including super resolution imaging, optical modulation, ultrafast switches, harmonic generation, colour filtering, and optical sensing.

# 1. Introduction

Metasurfaces, composed of arrays of artificially engineered subwavelength structures[1], have transformed photonic design, enabling compact and lightweight optical systems with unprecedented control over light[2]. Active metasurfaces made of phase-change materials (PCMs) provide reconfigurability[3], which has been extensively explored for tunable photonic devices, including optical switches[4], varifocal metalenses[5,6], spatial light modulators[7-10], photonic neural networks[11], and metafibres[12]. PCMs, including the germanium antimony telluride (GST) and germanium antimony selenide telluride (GSST) families[8, 12-14], vanadium dioxide ($VO_2$) [4, 15-18], and indium-antimony-telluride ($In_3SbTe_2$) [19-21], are typically selected for the development of active photonics in the mid- and long-wave infrared regimes. However, these PCMs inherently exhibit high optical absorption in the near-infrared (NIR) range, which hinders their use in resonant metasurfaces with high quality factors (Q-factors).

Photonic applications such as photoluminescence (PL) enhancement[22-28], harmonic generation[29-34], sensing[35-38], lasing[39-42], and quantum optics[43, 44] typically require strong resonances with high-Q factors. Recently, antimony sulfide ($Sb_2S_3$) and selenide ($Sb_2Se_3$) have emerged as a new PCM family exhibiting both high refractive index contrast and ultralow loss in the NIR region[45, 46]. They have been used for the development of tunable metalenses[5] and reconfigurable integrated photonic devices [47-49]. Meanwhile, bound states in the continuum (BICs), first proposed in quantum mechanics[50] as localized states with the energies embedded within the continuous spectrum of propagating waves, have been exploited for developing resonant metasurfaces with theoretically infinitely large Q-factors[42, 51]. Tunable BIC metasurfaces have been demonstrated by combining a thin (30-nm-thick) layer of $Sb_2S_3$ PCM with silicon-based BIC metasurfaces[52], or recently, a relatively thick (200-nm-thick) layer of conductive polymer with titanium-dioxide-based BIC metasurfaces[53]. Nevertheless, this combination of a thin layer of PCM with an existing metasurface made of a foreign material significantly limits the resonance tuning range to only a few nanometers, further complicated by a multi-step fabrication process.

Here we demonstrate the design and fabrication of tunable BIC metasurfaces directly utilising an ultralow-loss (25-nm-thick) $Sb_2Se_3$ thin film, patterned into an array of asymmetric double-rod structures (Fig. 1a). The length difference (asymmetry factor defined in Fig. 2c) between the double nanorods provides a convenient mechanism to control the radiative loss of the BIC metasurfaces[34], thereby supporting quasi-BIC (q-BIC) resonances accessible from far-field excitation. Benefiting from its high refractive index and ultralow loss in the NIR region (Fig. S1), we show that $Sb_2Se_3$ offers significant advantages over the silicon material typically[54] used for BIC metasurfaces in the NIR (Fig. 1b). The q-BIC resonance with high Q-factors and large modulation depths is supported by the $Sb_2Se_3$ metasurface even with a thickness down to 15 nm. In contrast, owing to a relatively low refractive index[55], the silicon metasurface counterpart cannot support a measurable q-BIC resonance when the thickness is reduced below 30 nm (Fig. S1c). Notably, ultrathin thickness is essential for PCMs for superior film amorphisation and phase-transition properties, enabling more effective heat transport and reduced power consumption[56].

Strikingly, the ultrathin $Sb_2Se_3$ BIC metasurfaces enable broad resonance tuning for the q-BIC resonance through the phase transition from amorphous to crystalline states (Fig. 1c). Broad resonance tuning over 20 nm and 170 nm can be achieved by a metasurface thickness of 25 nm and 150 nm, respectively. We experimentally fabricated 25-nm-thick BIC metasurfaces in $Sb_2Se_3$ with three stabilised states locked via annealing temperature during the phase transition, achieving a high Q-factor of ~130 and a large spectral tuning range of 16 nm in the NIR region. We demonstrate that fine-tuning the q-BIC resonance can match the peak absorption wavelength (~980 nm) of upconversion nanoparticles (UCNPs) [57-59], leading to a reduced excitation power for multiphoton photoluminescence and enhanced polarisation control of PL emission (Fig. 1d). This work offers a promising platform for developing active resonant metasurfaces in the NIR region, with significant implications in the fields of PL enhancement [60, 61], harmonic generation [29, 33], optical modulation [7-10], ultrafast switches[62, 63], colour filtering[64, 65], and optical sensing [35, 36].

## 2. Broad resonance tuning in high Q-factor Sb$_2$Se$_3$ metasurfaces

The unit cell of the Sb$_2$Se$_3$ metasurfaces comprises a double-rod structure positioned on a transparent silica substrate (Fig. 2a). We consider a rectangle-shaped unit cell with periodicities of $p_x$=410 nm and $p_y$=430 nm in the transverse plane. Without loss of generality, both rods have a fixed width of $W$=100 nm, and the separation distance between the two rods is $D$=200 nm. One of the nanorods has a fixed length of $L_1$=310 nm and the other nanorod has a length of $L_2 = L_1 + 2\Delta L$. Here, $\Delta L$ provides a convenient means to break the in-plane symmetry of the unit cell, controlling the far-field radiative loss of the structures and thereby the Q-factor of q-BIC metasurfaces. The asymmetry factor is defined as: $\alpha = \Delta L/L_2$. In our simulation, we consider plane-wave illumination with linear polarisation parallel to the long axis (x-axis) of the double-rod unit cell. We utilise experimentally measured refractive index data for the Sb$_2$Se$_3$ metasurface simulation (as shown in Fig. S1b). Notably, the ultralow-loss of Sb$_2$Se$_3$ in both amorphous and crystalline states enables the realisation of tunable BIC resonances. Without this low loss, strong resonances would easily dissipate.

We recognise that achieving ultrathin thickness is typically crucial for phase-change transitions in PCMs [56]. Simultaneously, optimising efficient light confinement and enhancing diploe moment interactions necessitates a specific thickness of high-index dielectrics. To explore this trade-off, we investigated the tuning range of q-BIC resonance in $Sb_2Se_3$ metasurfaces with varying thicknesses, ranging from 20 nm to 150 nm. The tuning range is defined as the difference in resonance wavelength between the crystalline and amorphous phase states ($\Delta\lambda_r = \lambda_c - \lambda_a$). In Fig. 2b, we demonstrate that a significant tuning bandwidth exceeding 175 nm can be achieved with a metasurface thickness of 150 nm, owing to the substantial refractive index contrast between the amorphous and crystalline states. The tuning range is notably sensitive to variations in thickness. As the thickness decreases, the tuning range reduces, accompanied by a blue shift of resonance down to around 1000 nm. Importantly, an ultrathin $Sb_2Se_3$ metasurface (25 nm) provides a generous tuning range of over 20 nm, making it particularly well-suited for PCM applications.

To verify the symmetry-protected q-BIC resonance, we studied the relationship between the resonance's Q-factors and the asymmetry factor $\alpha$ in a 25-nm-thick $Sb_2Se_3$ metasurface. By reducing the parameter $\Delta L$, we observed blue shifts and increased Q-factors across three selected annealing temperatures: $T_1$ (180 °C, mix-state), $T_2$ (200 °C, near-crystalline mix-state), and $T_3$ (220 °C, fully crystalline state) (see Fig. S2). Notably, at the $T_2$ state, we established a linear relationship: $Q \propto \alpha^{-2}$ in the logarithmic plot (Fig. 2c), providing strong evidence for the q-BIC nature of our resonance[51, 66]. The calculated Q factors at the three annealing temperatures consistently exceed 200 for varying parameter $\Delta L$ (Fig. 2d). It is worth noting that small changes in Q-factors occur during phase transitions due to negligible loss across the amorphous and crystalline states.

## 3. Fabrication and characterisation of $Sb_2Se_3$ metasurfaces

We fabricate 25-nm-thick $Sb_2Se_3$ metasurfaces using a standard planar lithography approach, which includes electron-beam lithography, lift-off, material deposition, and annealing processes (Fig. S3). The phase transition was achieved through a thermal annealing process, as schematically shown in Fig. 3a. The as-prepared metasurfaces were annealed at three different temperatures ($T_1$=180°C, $T_2$=200°C, and $T_3$=220°C), which are the same temperatures used in our simulation shown in Fig. 2. Fig. 3b presents optical and scanning electron microscope (SEM) images of the fabricated metasurface sample. The metasurface thickness of approximately 25 nm was confirmed using the tilted SEM image (Fig. S4) and the AFM measurement (Fig. 3c). The transmission spectra of the metasurfaces were characterised using a home-built bright-field imaging setup (see Methods).

We first demonstrate broad tuning of q-BIC resonance using a scaling factor ($S$), which adjusts the transverse dimensions of the metasurface unit cell by an overall multiplication factor of $S$. Spectral results are presented in Fig. 3d for three different scaling factors, ranging from $S$=1.6 to $S$=1.8. Thanks to the phase-change properties of $Sb_2Se_3$ metasurfaces, we can achieve precise resonance tuning post-fabrication. To demonstrate this principle, we experimentally locked the $Sb_2Se_3$ phases at three different temperatures, resulting in the resonance tuning shown in Fig. 3d. The resonance wavelength red shifts with increasing annealing temperature, matching well with the simulation results in Fig. S2. A large tuning range of 16 nm has been achieved for our 25-nm-thick $Sb_2Se_3$ metasurface. To the best of our knowledge, this is two times greater than previous efforts that combined a thin layer of PCM with an existing metasurface made of a different material[52]. Furthermore, we experimentally verified this symmetry-protected q-BIC resonance by varying the parameter $\Delta L$ (Fig. S5). The Q-factors were examined as a function of the asymmetry factor $\alpha$ in Fig. 3e. The linear relationship ($Q \propto \alpha^{-2}$) in the logarithmic plot in Fig. 3e confirms the q-BIC resonance, reaching a maximum Q-factor of over 100. Meanwhile, the Q-factors remained above 60 as $S$ varied from 1.6 to 2.0 (see Fig. 3f). We note that the drop of Q-factors at $S$=1.5 is related to the increased loss of the $Sb_2Se_3$ material at shorter NIR wavelengths. The original measurement spectra are provided in Fig. S5. Additionally, we experimentally examined the polarisation sensitivity of our $Sb_2Se_3$ metasurfaces. As predicted by theory, the modulation depths of our BIC metasurfaces should be maximised when the polarisation axis is aligned with the long axis of the double-rod structure, as symmetry breaking only occurs in this direction. Using a linear polariser in the excitation path, we recorded the normalised modulation depths at different polarisation angles, following a sinusoidal function as expected (Fig. 3g, refer to Fig. S6 for original spectra).

## 4. Modulating upconversion photoluminescence with $Sb_2Se_3$ metasurfaces

We used our developed $Sb_2Se_3$ metasurfaces to modulate PL in UCNPs with a peak absorption wavelength around 980 nm [57, 58]. We drop-casted a diluted $NaYF_4$:60%Yb, 2%Er, 4%Tm UCNPs solution onto the metasurfaces (Fig. 4a) and allowed it to dry naturally under ambient conditions. The UCNPs have an average size of approximately 45 nm (Fig. 4b), as shown in the statistical size distribution histogram (Fig. 4c). To investigate the modulation of upconversion PL with $Sb_2Se_3$ metasurfaces, we illuminated the metasurface sample with a femtosecond pulsed laser at a wavelength of 980 nm using a back-focal excitation configuration. More details about the optical setup are provided in the Methods section and Fig. S8.

Our UCNPs exhibit multiple emission bands with two peak wavelengths centred at 450 nm (peak 1) and 800 nm (peak 2). Fig. 4d presents the PL emission spectra under different excitation power densities, normalised to the maximum peak value at the highest excitation power we used. Upconversion emission involves a competitive process where different emission events contend for dominance, influenced by factors like excitation power density and energy transfer efficiency[59]. At lower power densities, the peak-2 emission dominates the spectra, while the peak-1 emission prevails at higher power densities[60]. We collected PL spectra from multiple UCNPs both on- and off-metasurfaces (The off-metasurface results are shown in Figure S7). We compared their peak intensity ratios (peak 1/peak 2) by increasing excitation power, exhibiting a saturated ratio of ~3.5 for both cases of UCNPs on- and off-metasurfaces (Fig. 4e). Notably, for the UCNPs on-metasurfaces, they exhibit approximately 4.5 times faster saturation than for their counterparts off-metasurface, suggesting reduced power for multiphoton photoluminescence.

Lastly, we illustrate the polarisation modulation in the PL emission by our metasurfaces, despite their mere 25 nm thickness. A linear polariser was incorporated in the detection optical path to examine the polarisation states of the collected PL spectra from UCNPs on- and off-metasurfaces. Given that the q-BIC resonance in $Sb_2Se_3$ metasurfaces is dependent on linear polarisation (as shown in Fig. 3g), it is expected that the PL emission modulated by the metasurfaces would display some polarisation characteristics. We explored the polarisation contrast in the PL emission between the component aligned with the long axis of the double-rod structure of $Sb_2Se_3$ metasurfaces and its perpendicular counterpart. We observed an approximate 30% contrast in PL intensity for the UCNPs on-metasurfaces (left panel in Fig. 4f), whereas their counterparts off-metasurfaces exhibited nearly identical responses (right panel in Fig. 4f). Figure 4g presents the integrated intensity of PL spectra as a function of the polariser rotation angle for UCNPs on- (red) and off-metasurfaces (gray). We found that the UCNPs on-metasurfaces exhibit some polarisation anisotropy, while their counterparts off-metasurfaces show an isotropic polarisation emission pattern.

## 5. Conclusion

We have demonstrated the design and fabrication of tunable BIC metasurfaces using ultralow-loss PCM $Sb_2Se_3$. Our fabricated BIC metasurfaces, with a thickness of only 25 nm, exhibit high Q-factors up to 130 and broad resonance tuning (16 nm) through the phase transition from amorphous to crystalline states. This work highlights the strong potential of $Sb_2Se_3$, which features a high refractive index and ultralow loss in the NIR region, for developing tunable nonlocal metasurfaces that support strong resonances. This approach offers significant advantages over previous experimental efforts that combined a thin layer of PCM with existing BIC metasurfaces made of different materials, which showed a significantly limited resonance tuning range and was further complicated by a multi-step fabrication process. In our future work, we will investigate the modulation of PCM $Sb_2Se_3$ using a pulsed laser beam [48, 49], aiming to achieve active resonance tuning in BIC metasurfaces. Experimentally, we employed these $Sb_2Se_3$ metasurfaces to modulate PL in UCNPs, reducing the excitation power for multiphoton PL and enabling emission polarisation manipulation. This work provides a promising platform for developing active resonant metasurfaces in the NIR regime, with broad applications including super resolution imaging[67, 68], optical modulation[7-10], ultrafast switches[62, 63], harmonic generation[29, 33], colour filtering[64, 65], and optical sensing[35, 36].

## 6. Methods

<u>Numerical simulations of BIC metasurfaces.</u> The numerically simulated transmission spectra of the $Sb_2Se_3$ metasurfaces were calculated using our in-house rigorously coupled wave analysis (RCWA) method. The substrate was considered as lossless quartz material with a fixed refractive index of 1.46. The metasurface structure on the substrate was set up with a periodic boundary condition and excited by transverse electric (TE) incidence parallel to the long axis of the double-rod structure. The optical properties, including the real and imaginary refractive indices of the $Sb_2Se_3$ material prepared under various temperature conditions, were provided in Fig. S1. For the simulation of the near-crystalline intermediate states $T_1$ and $T_2$, the imaginary part is set to be the same as the fully crystalline state $T_3$. The electric field distributions shown in Fig. 1b were performed using COMSOL Multiphysics.

<u>Fabrication of BIC Metasurfaces.</u> Fig. S3 illustrates the fabrication workflow of the BIC metasurfaces. First, a precleaned quartz substrate was spin-coated PMMA (A6) and form 500-nm-thick resist layer, followed by a soft baking process at 180 ºC. 10 nm Cr layer was sputtered onto the PMMA film to ensure the surface conductivity subsequent electron beam lithography (EBL). The Cr film was removed using a chromium etcher before the development. The EBL pattern was transferred to the PMMA layer after developing for 60 seconds in a MIBK:IPA (1:3) mixture. Finally, the samples were loaded into a vacuum chamber for the PCM deposition process.

<u>Growth of $Sb_2Se_3$ thin films.</u> $Sb_2Se_3$ thin films were deposited using a 99.99% pure target (Testbourne Ltd.). The deposition process involved 20 minutes of magnetron sputter deposition at 70 W RF power in an Ar environment (Kurt J Lesker), resulting in the formation of a 25 nm amorphous $Sb_2Se_3$ layer on the entire substrate. The amorphous BIC metasurface sample was then obtained by removing the unexposed PMMA resist using acetone through a lift-off technique. Three different crystalline phases of $Sb_2Se_3$ thin films were achieved through a controlled annealing process (linkam), with temperature conditions ranging from 180 °C to 220 °C while maintaining a constant annealing time of 8 minutes.

Synthesis of UCNPs and Morphological Characterisation. Raw materials, including $YCl_3·6H_2O$ (99.99%), $YbCl_3·6H_2O$ (99.9%), $ErCl_3·6H_2O$ (99.9%), $TmCl_3·6H_2O$ (99.99%), NaOH (98%), $NH_4F$ (99.99%), oleic acid (OA, 90%), and 1-octadecene (ODE, 90%), were purchased from Sigma-Aldrich. All chemicals were used as received without further purification unless otherwise noted. The $NaYF_4$:60%Yb,2%Er, and 4%Tm core nanoparticles were synthesized using a co-precipitation method.

In a typical procedure, a methanol solution containing 0.34 mmol $YCl_3·6H_2O$, 0.6 mmol $YbCl_3·6H_2O$, 0.02 mmol $ErCl_3·6H_2O$ and 0.04 mmol $TmCl_3·6H_2O$ was mixed with 6 mL oleic acid and 15mL 1-octadecene in a 50 ml three-neck round-bottom flask under vigorous stirring. The lanthanide oleate complexes were formed by heating the mixture to 150 °C for 30 minutes. The solution was then cooled to room temperature. Subsequently, a 10 mL methanol solution containing NaOH (2.5 mmol, 100 mg) and $NH_4F$ (4 mmol, 148 mg) was added and stirred at 50 °C for 30 minutes. The mixture was then slowly heated to 150°C and maintained for 40 minutes under argon flow to remove methanol and residual water. Next, the solution was quickly heated to 300 °C under argon flow for 1.5 hours before cooling to room temperature.

The resulting nanoparticles were precipitated by adding ethanol and collected by centrifugation at 9000 rpm for 5 minutes. After several washes with cyclohexane/ethanol/methanol, the final $NaYF_4$:60%Yb,2%Er,4%Tm nanoparticles were redispersed in cyclohexane at a concentration of 5 mg/mL for further experiments. The morphology of the UCNPs was characterised using a JEOL JEM F200 transmission electron microscope (TEM) operating at 200 kV. A dilute suspension of UCNPs dispersed in cyclohexane was dropped onto carbon-coated copper grids for imaging.

Optical characterisation of the $Sb_2Se_3$ metasurfaces and upconversion emission. We used a custom-built optical setup (see Fig. S8) for both the spectral characterisation of $Sb_2Se_3$ metasurfaces and PL measurements. A halogen lamp served as the light source for spectral measurements. A linear polariser and a half-wave plate were employed to ensure precise control of the linear polarisation states incident on the metasurfaces. For the upconversion PL measurements, we coupled a femtosecond pulsed laser at a wavelength of 980 nm into our custom-built optical setup and used an 842 nm short-pass filter to exclude the excitation wavelength, allowing only the PL signals to be collected. A second linear polariser was placed in the PL collection path to measure the polarisation dependency of the PL emission.

**Acknowledgements**

H.R. acknowledges funding support from the Australian Research Council (DE220101085, DP220102152). S.A.M. acknowledges funding support from the Australian Research Council (DP220102152) and the Lee Lucas Chair in Physics. K. M. and M. B. acknowledge project funding from the Australian Research Council through CE200100010. This work was performed in part at the Melbourne Centre for Nanofabrication (MCN) and at the Micro Nano Research Facility at RMIT University —both are Victorian Nodes of the Australian National Fabrication Facility (ANFF).

**Author contributions**

Z. X., C. Li, M. B., and H. R. proposed the concept and conceived the experiment; Z. X. performed the calculation in support from C. Li; Z. X., K. M., and C. Li designed and fabricated the metasurface samples; Z. X., H. Y., C. Li, and H. R. constructed the experiment, acquired the data and carried out the data analysis. S. A. M., M. B., and H. R. supervised the project; C. Liu, Y. L., and S. A. M. contributed to the discussions of the experimental results. Z. X., C. Li, and H.R. completed the writing of the paper with contributions from all the authors.

**Competing interests**: Authors declare that they have no competing interests.

**Data and materials availability**: All data are available in the main text or the supplementary materials.

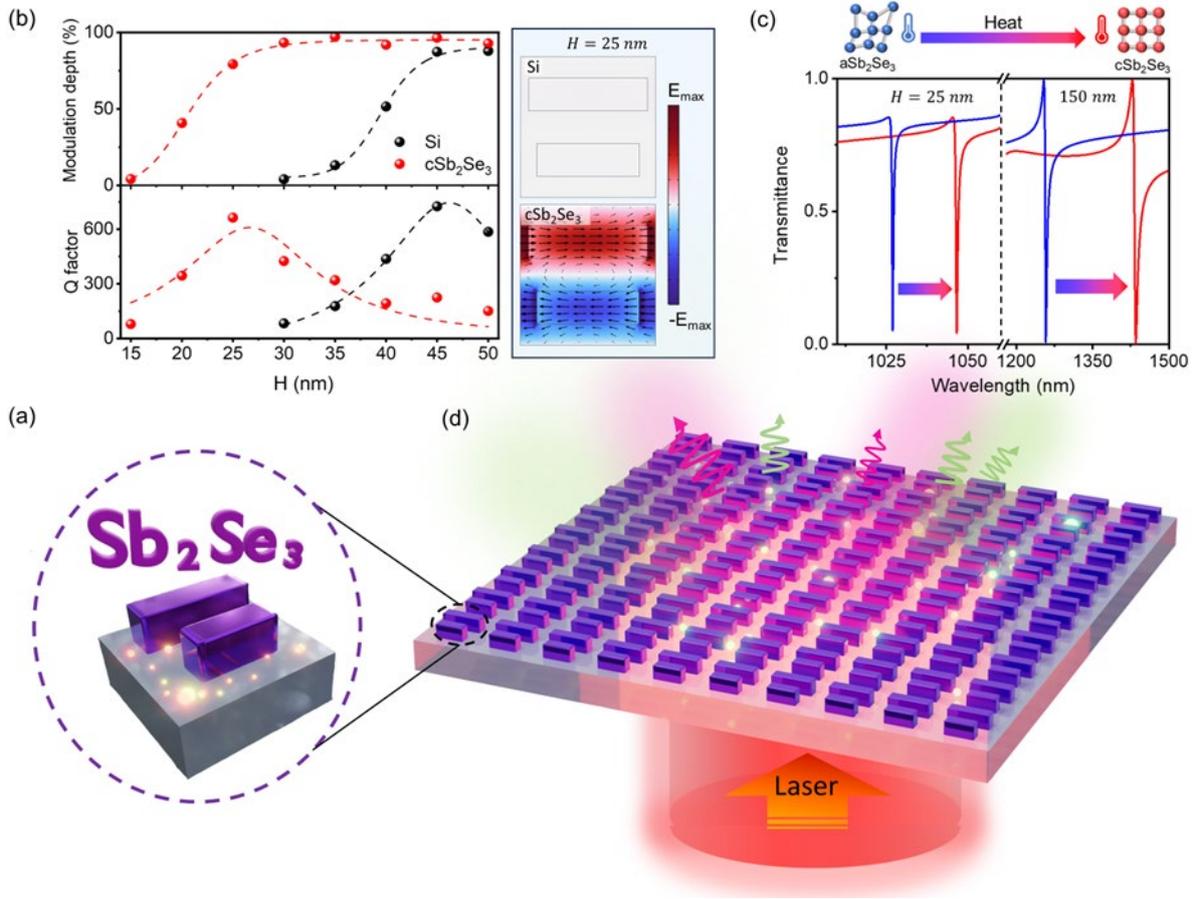

**Fig 1. Design of a 25-nm-thick BIC metasurface in low-loss Sb$_2$Se$_3$ for broad resonance tuning and PL modulation in UCNPs.** (**a**) Schematic of the unit cell structure of the Sb$_2$Se$_3$ BIC metasurface. (**b**) Comparison of simulated modulation depth ($\frac{T_{plateau}-T_{dip}}{T_{plateau}}$, where $T_{plateau}$ and $T_{dip}$ denote the plateau and resonance-dip intensities in the transmission spectra, respectively) and Q-factors between Sb$_2$Se$_3$ and Si metasurfaces composed of double-rod structures. Right panel: Field distributions of the silicon (top) and Sb$_2$Se$_3$ (bottom) unit cells with a thickness (*H*) of 25 nm. At the q-BIC resonance, the double-rod structure forms two anti-parallel in-plane electric field dipoles with slightly different amplitudes, canceling the far-field radiation due to destructive interference and resulting in high Q-factor resonance. (**c**) Broad resonance tuning is achieved in the transmission spectrum from Sb$_2$Se$_3$ metasurfaces with thicknesses of 25 nm (left) and 150 nm (right), based on the phase transition between the amorphous and crystalline states. The transmission dip is due to the cancelled far-field radiation at q-BIC resonance. (**d**) Schematic showing that upconversion nanoparticles are placed on the metasurface and pumped by a laser excitation, leading to multiphoton processes.

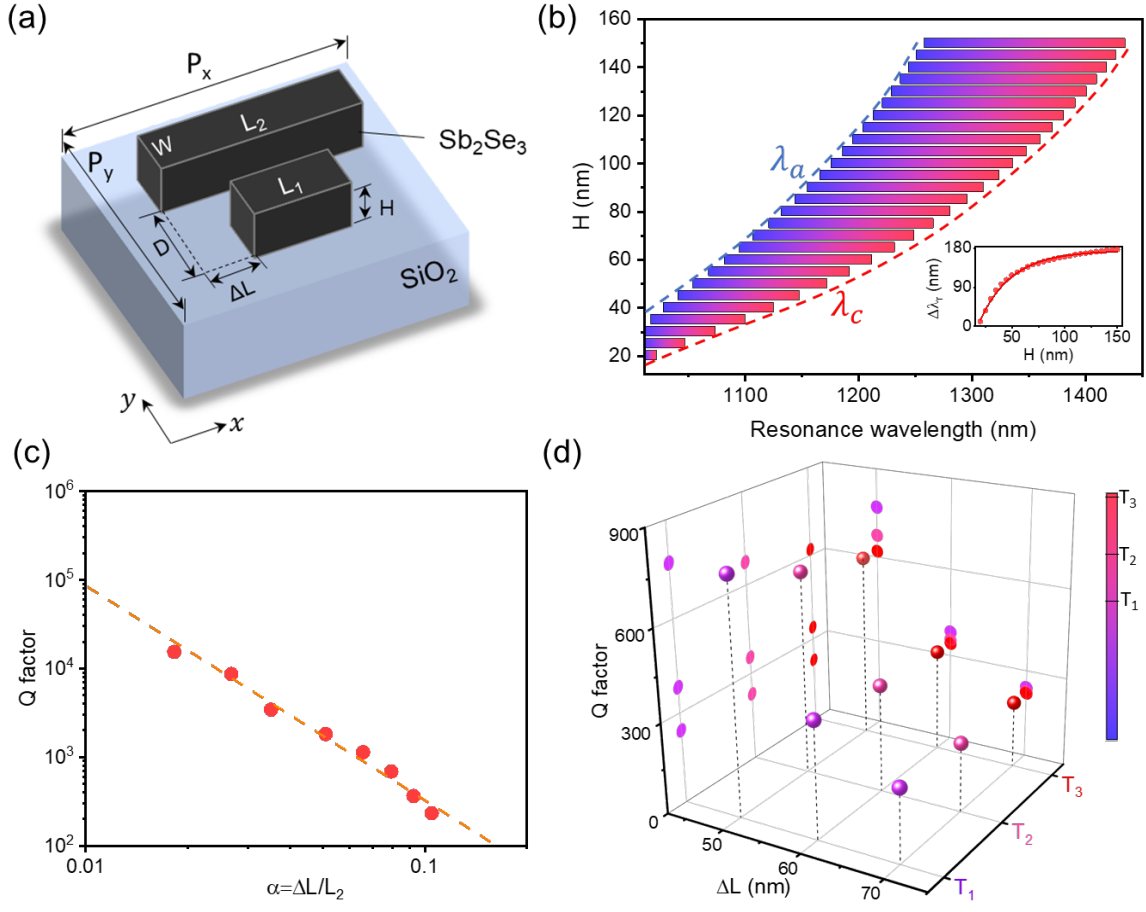

**Fig. 2. Broad resonance tuning in Sb$_2$Se$_3$ metasurfaces utilising symmetry-protected q-BIC resonance with high Q-factors.** (**a**) Schematic of the metasurface unit cell, labelled with size parameters. (**b**) Broad resonance tuning based on Sb$_2$Se$_3$ metasurfaces of varying thicknesses as the phase gradually transits from amorphous to crystalline states. The dashed lines represent the corresponding resonance wavelengths $\lambda_a$ and $\lambda_c$ in the amorphous and crystalline states. The inset highlights the tuning range as a function of thickness. (**c**) Linear fit of Q-factors with the asymmetry factor $\alpha$ for a 25-nm-thick Sb$_2$Se$_3$ metasurface annealed at T$_2$. (**d**) Examples of Q-factors for the 25-nm-thick Sb$_2$Se$_3$ metasurface annealed at three different temperatures (T$_1$=180°C, T$_2$=200°C, and T$_3$=220°C) with three different $\Delta L$ values.

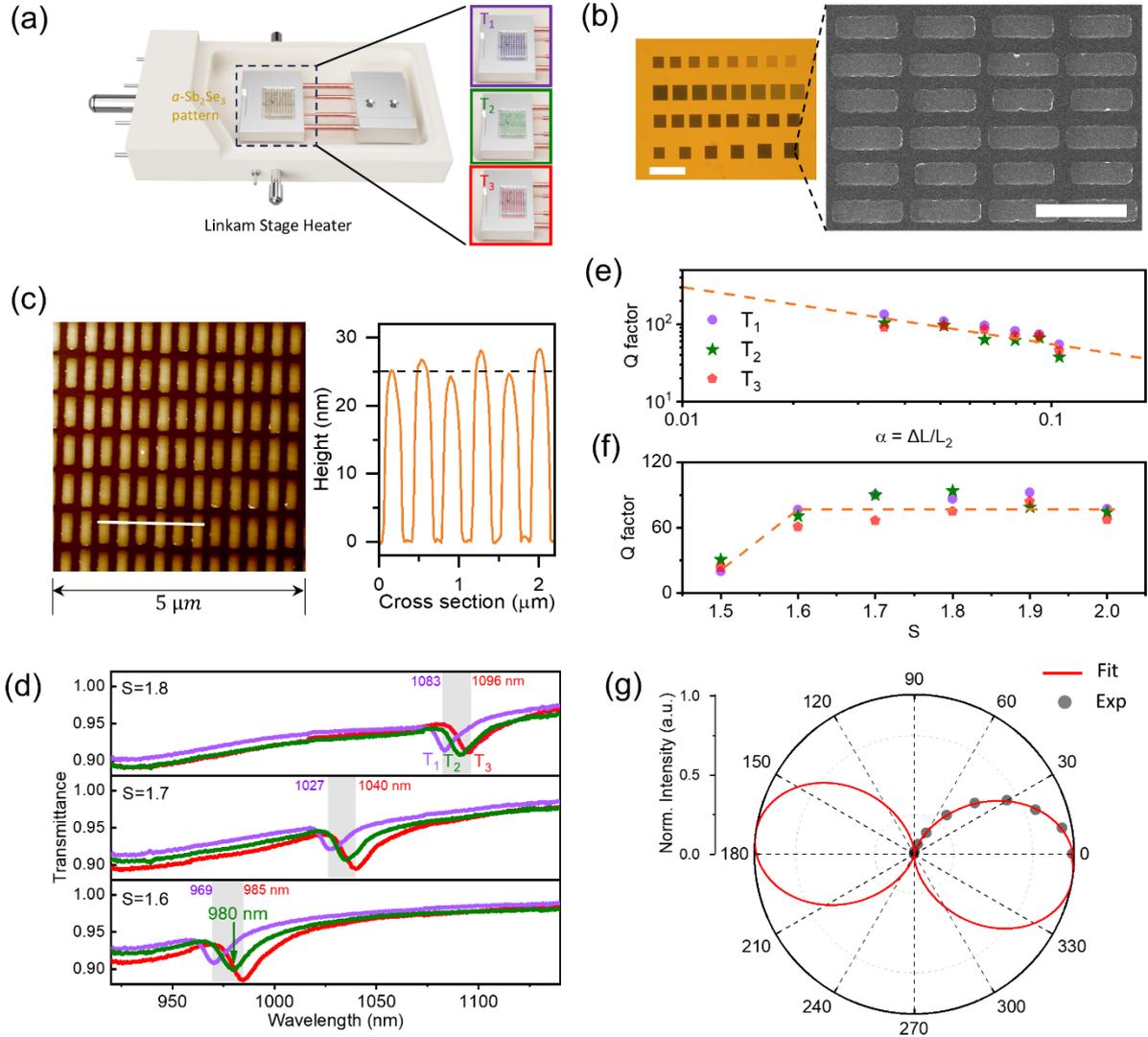

**Fig 3. Experimental characterisation of the fabricated 25-nm-thick $Sb_2Se_3$ metasurfaces.** (**a**) Schematic of the annealing setup for the $Sb_2Se_3$ metasurfaces where the insets present the temperatures $T_1 = 180 °C$, $T_2 = 200 °C$ and $T_3 = 220 °C$. (**b**) Optical and SEM images of the metasurface sample with scale bars of 200 μm and 1 μm, respectively. (**c**) AFM image of the $Sb_2Se_3$ structures. A cross section is cut by the white line in the image where corresponding height information is displayed in the right panel. (**d**) Transmission spectra of the $Sb_2Se_3$ metasurfaces annealed at different temperatures by varying the scaling factor $S$. (**e** and **f**) Measured Q-factors of the $Sb_2Se_3$ metasurfaces as functions of the asymmetric factor $\alpha$ (e) and $S$ (f). (**g**) The polarisation sensitivity of the $Sb_2Se_3$ metasurfaces supporting the symmetry-broken q-BIC resonance. This sample was fabricated at $T_2$ with $S = 1.7$ and $\Delta L = 70\ nm$.

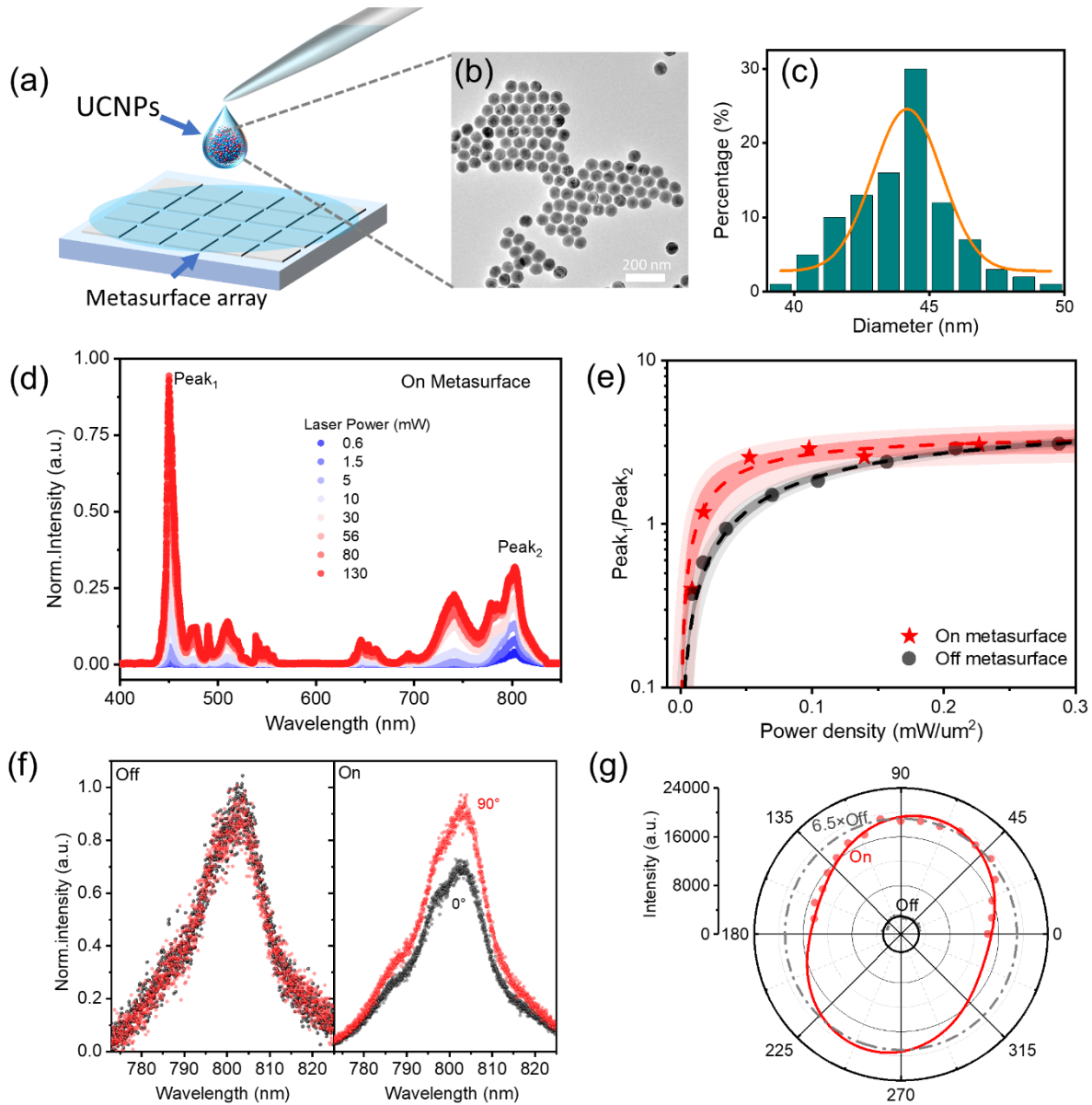

**Fig 4. Modulating upconversion PL with Sb$_2$Se$_3$ metasurfaces.** (**a**) Schematic of doping the UCNPs solution with a metasurface. (**b** and **c**) Transmission electron microscopy image (b) of the UCNPs and their size distribution histogram (c). The red curve shows the Gaussian fit to the data. (**d**) Normalised PL spectra of the UCNPs on-metasurfaces under different excitation laser powers. (**e**) The peak intensity ratios of PL emission as a function of excitation power density. Dashed curves are fitted results. (**f** and **g**) The polarisation analysis on the PL emission signals for UCNPs off- and on-metasurfaces. In (f), the PL spectra for UCNPs off- (left panel) and on-metasurfaces (right panel) are shown when the linear polariser's axis is aligned horizontally (black) and vertically (red). (**g**) Polar plots of the integrated emission for UCNPs on- (red) and off-metasurfaces (solid black). The dashed black curve shows a 6.5 times increased intensity compared to the solid black curve.